\documentclass{aa}  
\usepackage{graphicx}
\usepackage{txfonts}
\def\ms{\,ms$^{-1}$}         
\def\m2s2{\,m$^{2}$s$^{-2}$} 
\def\kms{\,kms$^{-1}$}       
\def\vsini{$v\sin{i}$}       
\def\Msol{M$_\odot$}         
\def\Rsol{R$_\odot$}

\begin{document}

\title{ELODIE metallicity-biased search for transiting Hot
  Jupiters\thanks{Based on observations collected with the ELODIE
    spectrograph on the 1.93-m telescope and with the 1.20-m telescope
    (Observatoire de Haute Provence, France)} } \subtitle{II. A
  very hot Jupiter transiting the bright K star HD\,189733}

\author{F. Bouchy\inst{1,2}, S. Udry\inst{3}, M. Mayor\inst{3}, C.
  Moutou\inst{1}, F. Pont\inst{3}, N. Iribarne\inst{2}, R. Da
  Silva\inst{3}, S. Ilovaisky\inst{2}, D. Queloz\inst{3},
  N.C.~Santos\inst{3,4}, D. Segransan\inst{3}, \and S. Zucker\inst{3,5}
  }

\offprints{\email{Francois.Bouchy@oamp.fr}}

\institute{Laboratoire d'Astrophysique de Marseille, 
               Traverse du Siphon, 13013 Marseille, France
\and
	   Observatoire de Haute Provence,
	      04870 St Michel l'Observatoire, France
\and
           Observatoire de Gen\`eve, 51 ch. des Maillettes, 1290
           Sauverny, Switzerland 
\and
	   Lisbon Observatory, Tapada da Ajuda, 1349-018 Lisboa, Portugal
\and
	   Weizmann Institute of Science, PO Box 26, Rehovot 76100, Israel
}

\date{Received ; accepted }
   
\authorrunning{Bouchy et al.}
\titlerunning{The transiting hot Jupiter HD\,189733\,b}

\abstract 
{Among the 160 known exoplanets, mainly detected in large
  radial-velocity surveys, only 8 have a characterization of their
  actual mass and radius thanks to the two complementary methods of
  detection: radial velocities and photometric transit.

We started in March 2004 an exoplanet-search programme biased toward
  high-metallicity stars which are more frequently host extra-solar planets.
  This survey aims to detect close-in giant
  planets, which are most likely to transit their host star.

For this programme, high-precision radial velocities are measured with
  the ELODIE fiber-fed spectrograph on the 1.93-m telescope, and
  high-precision photometry is obtained with the CCD Camera on the 1.20-m
  telescope, both at the Haute-Provence Observatory.

We report here the discovery of a new transiting hot Jupiter orbiting
  the star HD\,189733. The planetary nature of this object is
  confirmed by the observation of both the spectroscopic and
  photometric transits. The exoplanet HD\,189733\,b, with an orbital
  period of 2.219~days, has one of the shortest orbital periods
  detected by radial velocities, and presents the largest photometric
  depth in the light curve ($\sim3$\%) observed to date. We estimate
  for the planet a mass of $1.15$\,$\pm$\,0.04\,M$_J$ and a radius of
  1.26\,$\pm$\,0.03\,R$_J$. Considering that HD\,189733 has the same
  visual magnitude as the well known exoplanet host star HD\,209458, 
  further ground-based and
  space-based follow-up observations are very promising and will
  permit a characterization of the atmosphere and exosphere of this
  giant exoplanet.

}

\maketitle

\section{Introduction}

Within the last ten years more than 150 planetary systems have been
discovered, mostly by radial-velocity surveys. Although they provide a
great deal of information on the system orbital parameters and on the
host star properties, such surveys neither yield the acurate mass of the planet
(only $m$\,sin$i$) nor give any information about its size.  The
observation of planetary transits together with radial-velocity
measurements yield, on the other hand, the actual mass and planetary
radius (and thus mean density), providing constraints for
planet interior models.

Within the past two years, 6 exoplanets have been discovered first by
the photometric observation of a transit in front of the stellar disk
and then confirmed by spectroscopic follow-up: OGLE-TR-56b (Konacki et
al.  \cite{konacki03}), OGLE-TR-113b and 132b (Bouchy et al.
\cite{bouchy04}), TrES-1 (Alonso et al. \cite{alonso04}),
OGLE-TR-111b (Pont et al. \cite{pont04}) and OGLE-TR-10b (Bouchy et
al. \cite{bouchy05}; Konacki et al. \cite{konacki05}). Only 2
transiting exoplanets have been discovered first by radial velocities
and then had their photometric transit measured: HD209458 (Mazeh et
al. \cite{mazeh00}; Charbonneau et al. \cite{charbonneau00}; Henry et
al. \cite{henry00}) and HD149026 (Sato et al.  \cite{sato05}). The
host stars of the 5 OGLE planets are unfortunately faint and
complementary follow-up observations are difficult and time-consuming.
We present in this letter a new
transiting hot Jupiter orbiting the bright ($V$\,=\,7.7) and close
($d$\,=\,19\,pc) star HD\,189733.


\section{Observations}

HD\,189733 belongs to our ``ELODIE metallicity-biased search for
transiting Hot Jupiters'' survey (Da Silva et al. \cite{dasilva05}).
This survey started in March 2004 with the ELODIE spectrograph
(Baranne et al. \cite{baranne96}) on the 1.93-m telescope at the
Haute-Provence Observatory (France). The main idea of the programme is
to bias the target sample for high-metallicity stars which are more
likely to host planets (Gonzales \cite{gonzalez98}; Santos et al.
\cite{santos01}, \cite{santos05}; Fischer \& Valenti
\cite{fischer05}). It already allowed the discovery of 2
hot Jupiters orbiting the stars HD\,118203 and HD\,149143 (the latter
recently announced by Fischer et al. in prep).

The observational strategy of the survey is designed to primarily
target hot Jupiters which are ideal
candidates for follow-up photometric-transit searches. In practice,
the first spectroscopic measurement is made to estimate the
metallicity by measuring the surface of the cross-correlation function
of the ELODIE spectrum (Santos et al.  \cite{santos02}). Then, the
star is selected for further observations if the derived metallicity
[Fe/H] is greater than 0.1\,dex. The metallicity estimate requires a
spectrum with S/N$\ge$40.  For HD\,189733, this S/N ratio
was only reached after two independent exposures which, moreover,
revealed a large velocity variation.  The object was therefore
followed despite the fact that the measured metallicity was not larger
than 0.1.  Typical exposure times were between 15 and 25 minutes,
corresponding to photon-noise uncertainties of $\sim$\,5-7\,ms$^{-1}$.

A first set of 8 radial velocities of HD\,189733, showing a large-amplitude
variation, allowed us to easily constrain a circular orbital solution
with a very short period (2.2 days) for the companion.  With such a
short period, the probability that the companion crosses the stellar
disk is quite high ($\sim$1/8). We thus decided to attempt to measure
the transit both in spectroscopy (Rossiter-McLaughlin effect) using
the ELODIE spectrograph, and in photometry using the 1.20-m telescope
on the same site.  The f/6 Newton focus of the 1.20-m telescope is
equipped with a CCD camera system (1024$\times$1024 SITe
back-illuminated CCD) giving a field of view of 11.8 arcmin size and a
projected pixel size of 0.69 arcec.  A filter wheel holds the 
filters U', B, V, R$_c$ and I$_c$. The observation sampling was 
limited by the read-out time of the CCD controller (90\,s).  Basic data
processing and DAOPHOT stellar photometry (Stetson \cite{stetson87})
were applied to the images. Aperture photometry of
both brightest stars in the field was performed, with an aperture of
14 pixels. Correction of the sky background and cosmic removal were also
applied. The final light curves were then obtained using a single
reference star, and applying an extinction correction fitted on the data 
as a linear combination of the airmass.

\section{Stellar characteristics of HD189733}
\label{star}

HD\,189733 (HIP\,98505, GJ\,4130) is a dwarf star in the northern
hemisphere, listed in the Hipparcos catalog (ESA \cite{ESA97}) with a
visual magnitude $V$\,=\,7.67, a colour index $B-V$\,=\,0.932, and an
astrometric parallax $\pi$\,=\,51.94\,$\pm$\,0.87\,mas. This puts the
star at a close distance of 19.3\,pc from the Sun and allows us to
derive a corresponding absolute magnitude of $M_V$\,=\,6.25.  Although
the star is cataloged as G5, our analysis indicates a K1-K2 star. 
Following Santos et al. (\cite{santos04}), an
LTE high-resolution spectroscopic analysis of a high S/N spectrum
obtained with the CORALIE spectrograph gives T$_{\rm
  eff}$\,=\,5050\,$\pm$\,50\,K, $\log{g}$\,=\,4.53\,$\pm$\,0.14, and
[Fe/H]\,=\,$-$0.03\,$\pm$\,0.04.

\begin{table}
\caption{Parameters for the star HD\,189733, for the Keplerian
solution and inferred for the planetary companion.}            
\label{table1}      
\centering                         
\begin{tabular}{l l}        
\hline\hline                 
Period [days] & 2.219$\pm$0.0005 \\
Orbital eccentricity  & 0 [fixed] \\
Radial velocity semi-amplitude [\ms] & 205 $\pm$ 6 \\
Systemic velocity [\kms] & $-$2.361 $\pm$ 0.003 \\
O-C residuals [\ms] & 15 \\
& \\
Transit epoch [JD-2453000] & 629.3890 $\pm$ 0.0004 \\
Radius ratio & 0.172 $\pm$ 0.003\\
Impact parameter & 0.71 $\pm$ 0.02\\
Inclination angle [$^o$] & 85.3 $\pm$ 0.1 \\
& \\
Temperature [K] & 5050$\pm$50 \\
log\,$g$ & 4.53 $\pm$ 0.14 \\
$[\rm{Fe/H}]$ & -0.03 $\pm$ 0.04 \\
{\vsini} [\kms] & 3.5 $\pm$ 1.0 \\
Star mass [\Msol] & 0.82 $\pm$ 0.03 \\
Star radius [\Rsol] & 0.76 $\pm$ 0.01 \\
$P_{rot}$ [days]    & $\sim$\,11 \\
& \\
Orbital semi-major axis [AU] & 0.0313 $\pm$ 0.0004 \\
Planet mass [M$_J$ ]  & 1.15 $\pm$ 0.04 \\
Planet radius [R$_J$]  & 1.26 $\pm$ 0.03 \\
Planet density [$g\;cm^{-3}$] & 0.75 $\pm$ 0.08 \\
\hline                                  
\end{tabular}
\end{table}

The $V-K$ colour implies a temperature of 4996\,$\pm$\,40\,K with the
calibration of Kervella et al. (\cite{kervella04}), and the
Str\"omgren photometry gives 4950\,$\pm$\,150 K, both estimates
coherent with the spectroscopic temperature.  Confronting the
spectroscopic parameters obtained for HD\,189733 with the Girardi et
al. (\cite{girardi02}) stellar evolution models gives a radius of
0.761\,$\pm$\,0.014\,\Rsol, and a mass of 0.81\,$\pm$\,0.03\,{\Msol}
using the $V$ magnitude, or 0.82\,$\pm$\,0.025\,{\Msol} using the $K$
magnitude. From the Baraffe et al. (\cite{baraffe98}) models, we find
$R$\,=\,0.75\,$\pm$\,0.01\,{\Rsol} and $M$\,=\,0.80-0.85\,{\Msol}.
Kervella et al. (\cite{kervella04}) have calibrated a relation between the $V-K$
colour and radii measured from interferometry that shows very little
dispersion for low-mass stars ($\sim$\,1\,\%).  This calibration gives
a radius of 0.77\,{\Rsol} for HD\,189733.  Empirical and theoretical
estimates of the host star's radius are thus in excellent agreement,
with a small error interval, thanks to the star's position in a thin
and slowly-evolving part of the lower main sequence and to the
precision of the Hipparcos parallax. In the subsequent analysis we
combine the above estimates into $R$\,=\,0.76\,$\pm$\,0.01\,{\Rsol}
and $M$\,=\,0.82\,$\pm$\,0.03\,\Rsol.

A stellar rotation $v\sin{i}$\,=\,3.5\,$\pm$\,1.0\,{\kms} is estimated
from the calibration of the cross-correlation functions. Assuming that
the stellar spin axis is perpendicular to the line of sight, we can
derive the stellar rotational period of $\sim$11 days.  The
chromospheric activity index $S$ based on the relative flux level on
CaII H and K lines was measured by Wright et al. (\cite{wright04}).
The value of $S=0.525$ indicates a relatively active star.

Table\,\ref{table1} lists the observed and derived parameters of the
star HD\,189733.

\begin{figure}[t]
\includegraphics[width=8.5cm]{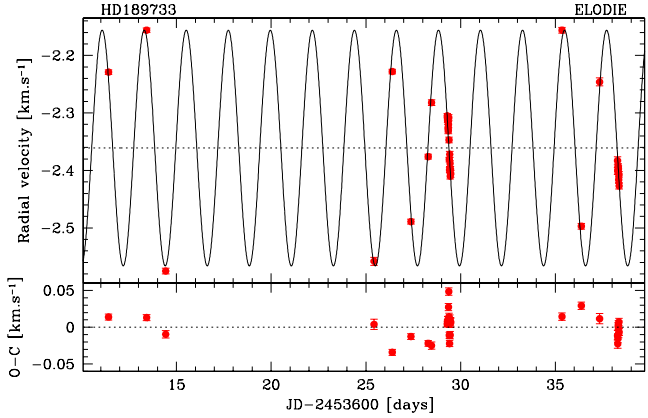}
\caption{Radial-velocity measurements of HD\,189733 superimposed on 
  the best Keplerian solution. The higher density of points
  corresponds to spectroscopic-transit measurements. They are
  not used to derive the Keplerian solution.  Error bars represent the
  photon-noise uncertainties. }
\label{rv}
\end{figure}
    
\section{Keplerian solution and spectroscopic transit}
\label{orbit}

\begin{figure}
\includegraphics[width=8.5cm]{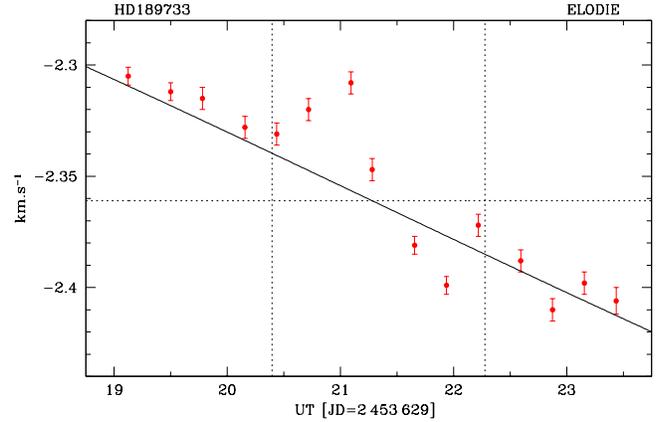}
\caption{Radial velocity sequence made on September 15th 2005 exhibiting the 
  Rossiter-McLaughlin effect. The vertical dashed lines correspond to
  the first contact and last contact of the transit deduced from the
  photometric light curve.}
\label{rvzoom}
\end{figure}

Radial velocity (RV) measurements of HD\,189733 were conducted in
August and September 2005 (from JD=2\,453\,611 to 2\,453\,638).
Figure\,\ref{rv} shows the RV measurements together with the derived
Keplerian solution. 
The short sequence of RV measurements made during the night
2\,453\,629 is displayed on Fig.\,\ref{rvzoom} and clearly shows the
RV anomaly due to the Rossiter-McLaughlin effect (spectroscopic
transit). A deviation from the Keplerian solution of about
$\pm$\,40\ms\ occurs because the transiting planet occults first the
approaching limb and then the receding limb of the rotating star. 
The observation of this effect provides an unambiguous
confirmation of the transiting planet.  Thanks to the on-line data
reduction of ELODIE spectra, the RV anomaly was observed in real time
during the night and revealed the transit of a planetary companion
before the photometric analysis.  HD\,189733 is the third star
known to present a Rossiter-McLaughlin effect due to a planetary companion
and actually the first to be identified as a transiting planet {\it 
spectroscopically}.
Like HD\,209458 (Queloz et al.  \cite{queloz00a}) and HD\,149026
(Sato et al. \cite{sato05}), HD\,189733 presents a positive RV
anomaly during the ingress phase of the transit and a negative RV
anomaly in the egress phase, indicating that the stellar rotation is
in the same direction as the planet motion. The symmetrical deviation
seems to occur at mid-transit, indicating that the orbital plane is
quite coplanar with the stellar equatorial plane.  The amplitude of
the anomaly is comparable to the one measured on HD209458 by Queloz et
al. (\cite{queloz00a}), in agreement with the fact that the star's
{\vsini} and the radius ratio between the planet and its host star are
very close for these two systems.

\begin{figure}
\includegraphics[width=8.5cm]{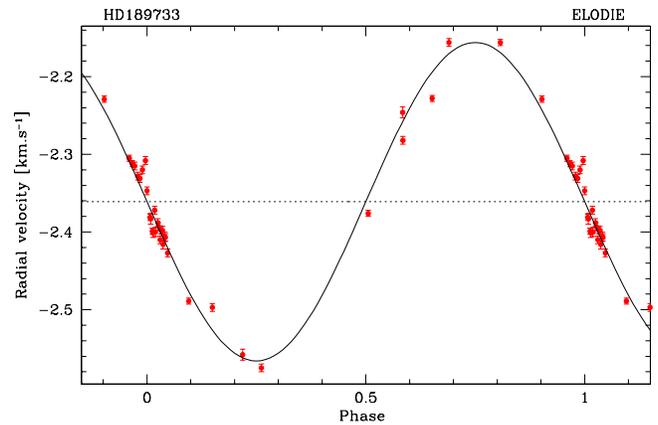}
\caption{Phase-folded radial velocity measurements of HD\,189733 
  superimposed on the best Keplerian solution. Error bars represent
  the photon-noise uncertainties.}
\label{phase}
\end{figure}

The Keplerian solution is derived without the RV points obtained
during the spectroscopic transit and using the constraint 
of transit epochs given by the observed photometric transits. 
The best fit to the data yields a short-period
orbit ($P$\,=\,2.219 d) with an eccentricity compatible with zero.  
The phase-folded radial-velocity
curve is displayed in Fig.\,\ref{phase}.  The orbital elements are
listed in Table\,1, jointly with the inferred stellar and planetary
parameters. The somewhat large residuals around the solution (15\ms)
are probably explained by the activity-induced jitter of the star.
We checked (on the cross-correlation functions) that the shape of the spectral lines was
not varying in phase with the radial-velocity change, what would be expected
in case of spot-induced RV variations.

\section{Photometric transit and characterisation of HD\,189733\,b}

\begin{figure}
\includegraphics[width=8.5cm]{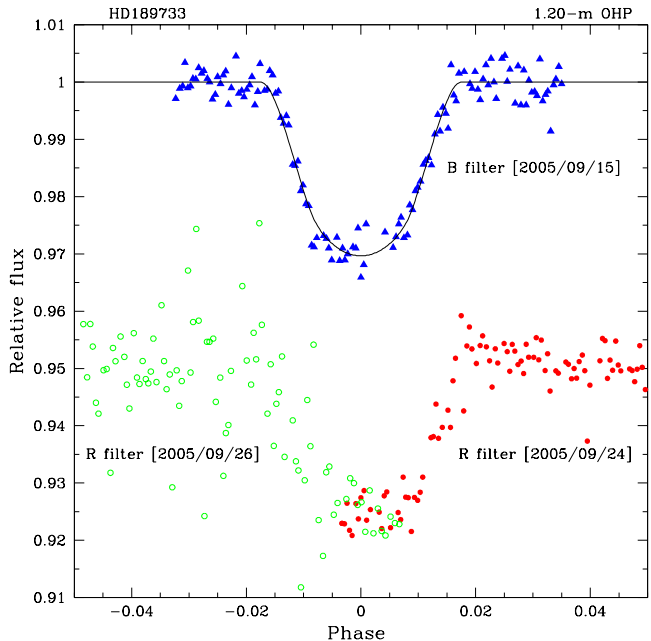}
\caption{Photometric transits of HD\,189733 observed with 
  the 1.20-m OHP telescope. Triangles correspond to the 
  observation on B band made on September 15th 2005. Full circles 
  and open circles correspond to observation on R band made respectively on 
  September 24th and 26th 2005. The solid curve represents the best-fit model 
  for the complet B-band transit.}
\label{photom}
\end{figure}

From the ephemeris predicted by the radial velocities, 3 transit
events (nights 2\,453\,629, 2\,453\,638 and 2\,453\,640) have been
followed in photometry with the 1.20-m telescope at OHP. The first
night, a complete photometric transit was observed in the B band
(Fig.\,\ref{photom}). For the next 2 attempts, performed in the R band, 
only partial coverage of the transit was possible. 
We observed the transit egress on the first night and the
ingress on the second. Because of non-optimum atmospheric conditions,
the partial transit measurements are of poorer quality than the B-band
observations, they are thus not used for our determination of the
planetary parameters. They are however of prime importance to confirm
the transit detection and precisely specify the orbital period.

The dispersion of the light curve in the B band (Fig.\,\ref{photom})
is about 2\,mmag at the beginning of the sequence and 3\,mmag at the
end.  The photon noise is 1.1\,mmag, and the total dispersion is
primarily due to the photon-noise on the comparison star and to the
increasing airmass.  A transit light curve was fitted to the data
using the mass and radius found for the host star in
Sect.\,\ref{star}, the orbital parameters of Sect.\,\ref{orbit} and
limb darkening coefficients in the B filter from Claret
(\cite{claret00}) for $T_{\rm eff}$\,=\,5000\,K, $\log{g}$\,=\,4.5 and
[M/H]\,=\,0.  The free parameters are the transit central epoch, the
radius ratio between the star and planet, and the inclination angle of
the orbit.  We find $T_{tr}$\,=\,2\,453\,629.3890\,$\pm$\,0.0004,
$R_{pl}/R$\,=\,0.172\,$\pm$\,0.003 and $i$\,=\,85.3\,$\pm$\,0.1.  The
formal uncertainties are very small. The dominant source of error is
likely to be the systematics in the photometry. To estimate their
effect, we repeated the reduction of the photometry using different
procedures and different comparison stars.  This resulted in a change
of 4\,\% for the radius ratio and 0.3$^{o}$ for the inclination angle.
The main parameters of the planet are therefore determined with
remarkable accuracy, even from the "discovery" data, thanks to the
very good determination of the primary star parameters. Orbital and
physical parameters of HD\,189733\,b are listed in
Table\,\ref{table1}.


Hipparcos (ESA \cite{ESA97}) observed HD189733 and obtained 176 reliable photometric
measurements. Following Soderhjelm (\cite{soderhjelm99}), Robichon \& Arenou 
(\cite{robichon00}) and Castellano et al. (\cite{castellano00}), we 
attempted to search for planetary transits in the Hipparcos data of
HD189733. We looked for a transit signal using the same shape and depth 
of the transit observed with the 1.20-m. Our best solution gives a $\chi^2$ of 
249 for 171 degrees of freedom. This quite large value is explained by the 0.03 mag 
variability of the star which was already mentioned in the Hipparcos catalog. 
Our best-fit period for the Hipparcos data is P\,=\,2.218575\,$\pm$ \,0.000003 days. 
This result was also found and is described in details 
in H\'ebrard \& Lecavelier des Etangs (\cite{hebrard05}).
  
\begin{figure}
\includegraphics[width=8.5cm]{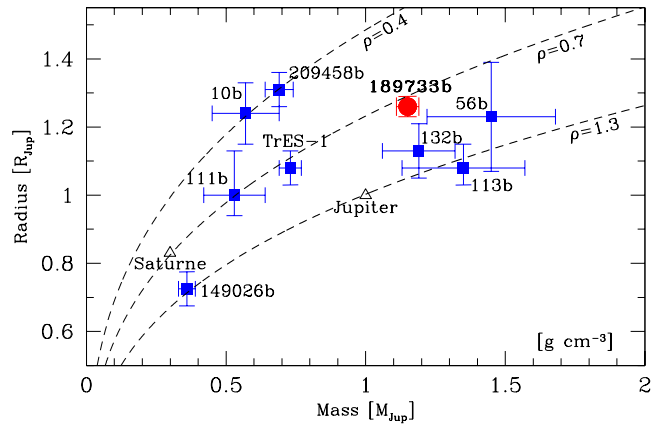}
\caption{Mass-radius diagram for the 9 transiting exoplanets. 
  Jupiter and Saturn are indicated for comparison, as well as the loci
  of isodensities at 0.3, 0.7 and 1.3\,g\,cm$^{-3}$. Data are from
  Pont et al. \cite{pont04} (OGLE-TR-111b), Konacki et al.
  \cite{konacki05} (OGLE-TR-10b), Moutou et al. \cite{moutou04}
  (OGLE-TR-132b), Torres et al.  \cite{torres04} (OGLE-TR-56b), Bouchy
  et al. \cite{bouchy04} (OGLE-TR-113b), Laughlin et al.
  \cite{laughlin05} (HD209458b and TrES-1), Sato et al. \cite{sato05}
  (HD149026b).}
\label{mr}
\end{figure}

\section{Summary and concluding remarks}

We have presented the characteristics of the new transiting hot
Jupiter in orbit around the star HD\,189733, detected by the new
planet-search programme conducted with the ELODIE spectrograph.  The
period derived from the RV measurements is very short ($P$\,=\,2.219\,d)
and the orbit is circular. The photometric transit measurements allow
the determination of the planetary mass (1.15\,$\pm$\,0.04\,M$_J$),
radius (1.26\,$\pm$\,0.03\,R$_J$) and mean density
(0.75\,$\pm$\,0.08\,g\,cm$^{-3}$).

Although our programme is biased towards metal-rich stars, the new
candidate orbits a solar-metallicity star. Its short period tends then
to weaken the proposed relation between separation and metallicity for hot
Jupiters (Queloz et al. \cite{queloz00b}; Sozzetti \cite{sozzetti04}).

Figure\,\ref{mr} presents the mass-radius diagram of the 9 known
transiting exoplanets.  In term of mass and radius, HD\,189733\,b is quite
similar to the very hot Jupiters OGLE-TR-56\,b, 113\,b and 132\,b.

\begin{figure}
\includegraphics[width=8.5cm]{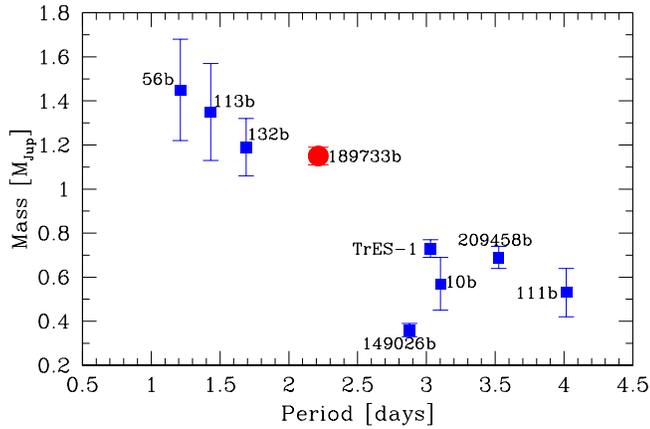}
\caption{Mass-period diagram for the 9 transiting exoplanets.}
\label{pm}
\end{figure}

Figure\,\ref{pm} displays the period-mass diagram of the 9
known transiting exoplanets. HD\,189733\,b appears to be intermediate
between the class of very hot Jupiters and hot Jupiters and provides in
some way the missing link between planets from transit and
radial-velocity surveys in terms of mass and period. HD\,189733\,b
confirms the correlation between the periods (or orbital distances)
and masses of transiting exoplanets pointed out by Mazeh et al.
(\cite{mazeh05}). At such a close distance from the star, it is likely that 
HD\,189733\,b undergoes some evaporation (Lecavelier et al. \cite{lecavelier04}; 
Baraffe et al. \cite{baraffe04}).

With the same visual magnitude as HD\,209458 and even brighter in
infrared, HD\,189733 belongs to the very short list of bright stars
with detected planetary transits. Only HD\,209458\,b, TrES-1,
HD\,149029\,b and HD\,189733\,b have parent stars brighter than
$V$\,=\,12. They therefore provide primary targets for additional
ground-based and space-based measurements requiring very high
signal-to-noise ratio observations.

\begin{acknowledgements}
  We are grateful to all the night assistants and
  telescope staff of Observatoire de Haute Provence for their efforts
  and their efficiency.  We wish to thank the Programme National de
  Planetologie (PNP), the Swiss National Science Foundation (FNRS)
  and the Geneva University for their continuous support to our
  planet-search programs. NCS would like to thank the support from
  Funda\c{c}\~ao para a Ci\^encia e a Tecnologia (Portugal) the form
  of a scholarship (reference SFRH/BPD/8116/2002) and a grant
  (reference POCI/CTE-AST/56453/2004).
\end{acknowledgements}

\end{document}